\documentstyle[12pt]{article}
\begin{document}
\title{Symmetry Properties of the $k$--Body Embedded Unitary
Gaussian Ensemble of Random Matrices\\}
\author{
\\ \\
Z. Pluha\v r\\
Charles University, Prague, Czech Republic\\ \\
and\\ \\
H.A. Weidenm\"uller\\
Max-Planck-Institut f\"ur Kernphysik, \\
D-69029 Heidelberg, Germany\\ \\ \\}
\date{\today}
\maketitle
\begin{abstract}
We extend the recent study of the $k$--body embedded Gaussian
ensembles by Benet {\it et al.} (Phys. Rev. Lett. 87 (2001) 101601-1
and Ann. Phys. 292 (2001) 67) and by Asaga {\it et al.}
(cond-mat/0107363 and cond-mat/0107364). We show that central results
of these papers can be derived directly from the symmetry properties
of both, the many--particle states and the random $k$--body
interaction. We offer new insight into the structure of the matrix of
second moments of the embedded ensemble, and of the supersymmetry
approach. We extend the concept of the embedded ensemble and define
it purely group--theoretically. 
\end{abstract}

\newpage

\section{Introduction}
\label{int}

Recently, novel results on the spectral properties of the $k$--body
embedded unitary and orthogonal ensembles of random matrices were
obtained both for Fermions (Refs.~\cite{ben00L} and \cite{ben00A})
and for Bosons (Refs.~\cite{asa01L} and \cite{asa01A}). We refer to
these two pairs of papers as to ABRW. These results were based on 
explicit analytical calculations. The authors did mention that their 
work had its root in the underlying symmetry properties of the 
embedded ensembles. They did not, however, display these symmetry 
properties explicitly, nor did they derive their results with the help 
of such properties.

In the present paper, we remedy this situation. We show that central
results of ABRW follow from the symmetry properties of the
many--particle states and of the random $k$--body interaction. For
simplicity, we confine ourselves to the $k$--body embedded unitary
ensemble EGUE($k$) although our results can certainly be generalized
to the orthogonal case.

To define EGUE($k$), we consider $m$ particles (Fermions or Bosons)
in $l$ degenerate single--particle states. The random $k$--body
interaction connects states in the Hilbert space of these $m$
particles. It is useful to introduce, for arbitrary $t$, the
dimension $N_t$ of the Hilbert space of $t$ particles. For Fermions,
we have $N_t = { l \choose t}$ while for Bosons, $N_t = { l + t - 1
\choose t}$. For simplicity, we write $N$ for $N_m$. We show that for
the embedded ensemble, three symmetry groups are relevant. These are
(i) the group SU($l$) of unitary transformations of the degenerate
$l$ single--particle states; (ii) the group U($N_k$) of unitary
transformations of the $k$--body interaction; (iii) the group U($N$)
of unitary transformations of the Hilbert space. The group SU($l$)
governs the embedding. The group U($N_k$) is the symmetry group of
EGUE($k$). The group U($N$) is the symmetry group of the Gaussian
unitary ensemble (GUE), i.e., the free Gaussian unitary ensemble in
$N$ dimensions.

The matrix elements of the $k$--body interaction in the space of
$m$--particle states are Gaussian random variables with zero mean.
Therefore, all information necessary for calculating averages over
EGUE($k$) is contained in the matrix of second moments of the
ensemble. We show that the spectral representation (the ``eigenvalue
expansion'' of ABRW) of the matrix of second moments constructed
explicitly in ABRW can also be obtained from symmetry arguments.
To this end, we decompose the random $k$--body interaction into a sum
of $k$--particle basic interaction terms which transform according to
the irreducible representations of SU($l$). The latter are specified
by the bodyness (unitary rank) quantum number $b$ introduced by Mon and
French~\cite{mon75}. The eigenvalues appearing in the spectral
representation of the matrix of second moments are uniquely defined by
this quantum number, and the eigenvectors are given by SU($l$)
Clebsch--Gordan coefficients.

We isolate the U($N$)--invariant part of the matrix of second moments.
The result is unexpected as this part consists of two pieces. We find
the expected term which has the form of the matrix of second moments
of the free GUE, or, equivalently, of EGUE($m$). However, we also find 
another term which has the form of the matrix of second moments of the
degenerate Gaussian ensemble of multiples of the unit matrix, or,
equivalently, of EGUE($0$). We show that for $k \ll m$, the
U($N$)--invariant part favors a Gaussian spectral shape and
non--ergodic spectral properties. For $2k > m$, on the other hand,
that part favors a semicircular spectral shape, and ergodic behavior.

We also investigate the invariance of the supersymmetric $n$-point 
generating functions of the ensemble under the unitary transformations
of the interactions, and we examine the consequences of that symmetry 
for the saddle--point solutions. Finally, we extend the concept of the 
embedded ensemble to cases where the embedding is governed by an 
arbitrary compact group different from SU($l$).

The paper is organized in the following way. Some basic relations for
the representations of SU($l$) in Fock space are presented in
Section~\ref{for}. The ensemble EGUE($k$), its symmetry and the SU($l$)
expansion of the embedded interactions are discussed in
Section~\ref{ens}. The symmetry properties of the matrix of second
moments are discussed in Section~\ref{mom}. In Section~\ref{inv}, we
isolate the U($N$)--invariant part of the matrix of second moments and
determine its relative importance. The properties of the supersymmetric
generating functions are investigated in Section~\ref{gen}. Comments
on the application of the ABRW approach to other ensembles and
concluding remarks are contained in Section~\ref{con}. Some
mathematical details, including the evaluation of eigenvalues of the
matrix of second moments, are given in Appendices. We always treat
Fermions and Bosons jointly. For the representation theory we follow
the books by Wybourne \cite{wyb74} and Lichtenberg \cite{lic78}.

\section{Fock Space and the Group SU($l$)}
\label{for}

In this Section, we introduce the group SU($l$) and describe the
action of group operations on composite operators. We use a running
label $i=1, \ldots, l$ for the degenerate single--particle states. We
use second quantization and denote the single--particle creation and
annihilation operators by $a_i^+ $ and $a_i$, respectively. The vacuum
state is written as $| \, \rangle$. We consider both Fermions and
Bosons. Hilbert space is spanned by $N = N_m$ orthonormal $m$-particle
states denoted by $|m \, v_m \rangle$. The index $v_m = 1, \ldots, N$
is a running label. We emphasize that the basis states $|m \, v_m
\rangle$ are not necessarily simple states (Slater determinants for
Fermions or product states for Bosons as used in ABRW): Linear
combinations of such states are also admitted. No specific assumption
on the choice of the basis will be made. The creation operators for the
basis states are denoted by $A^+(m v_m)$.

An element $u \in \mbox{SU}(l)$ generates a unitary transformation of
the single--particle states and a corresponding transformation 
\begin{equation}
T(u) a_i^+ T^+(u) = \sum_{i'} u_{i' i} a_{i'}^+ \; ,
\qquad u \in \mbox {SU}(l) \; ,
\label{TaT}
\end{equation}
of the creation operators $a_i^+$. The operators $T(u)$ are generated
by the traceless linear forms of $a_i^+a_{i'}$. The transformation $u$
induces a linear transformation of the operators $A^+(m v_m)$. These
operators transform according to the irreducible representations
$D^{f_m}(u)$,
\begin{equation}
T(u) A^+(m v_m) T^+(u)= \sum_{ v'_m }
D_{ v'_m v_m}^{f_m} (u) A^+(m v'_m ) \; .
\label{TAT}
\end{equation}
In the Fermionic (Bosonic) case, the matrices $D^{f_m}(u)$ are given,
to within an equivalence transformation, by the totally antisymmetrized
(totally symmetrized, respectively) powers of $u$. In the sequel, we
write the Young tableaux of the irreducible representations of SU($l$)
as $h=(h_1, \ldots, h_{l-1})$, with $h_j$ denoting the number of columns
of length $j$. The symbol $0^j$ stands for $j$ zeros. In the case of
Fermions, $D^{f_m}(u)$ belongs to the Young tableau $f_m =
(0^{m-1}10^{l-m-1})$; in the case of Bosons, it belongs to the Young
tableau $f_m = (m 0^{l-2})$. The annihilation operators $A(m v_m)$
transform according to the conjugate irreducible representation
${\cal D}^{ {\bar f}_m }(u) = ( D^{f_m}(u) )^\ast$, where ${\bar f}_m =
(0^{l-m-1} 1 0^{m-1})$ for Fermions, and ${\bar f}_m = (0^{l-2}m)$ for
Bosons.

In analogy with the familiar fractional--parentage technique of first
quantization, we can expand the $m$--particle states $|m v_m \,
\rangle$ into a sum of products of the $k$-particle creation operators
$A^+(k v_k)$ and the $(m-k)$-particle states $|m\!-\!k \, v_{m-k} \,
\rangle$. Explicitly,
\begin{equation}
|m\,v_m \, \rangle
= {m \choose k}^{-1/2}
\sum_{v_k v_s} A^+(k v_k ) |s v_s \, \rangle
C_{ f_k v_k f_s v_s }^{ f_m v_m } \; .
\label{cfp}
\end{equation}
Because of the duality relation, Eq.~(\ref{mom-dua}) below, the
particle rank $k$ and the particle rank $m - k$ are intimately linked.
This is why we use here and in the sequel the label
\begin{equation}
s = m - k.
\label{s}
\end{equation}
The coefficients $C_{ f_k v_k f_s v_s }^{ f_m v_m }$ are the SU($l$)
Clebsch--Gordan coefficients for the coupling of basic states of the
represenations $D^{f_k}(u)$ and $D^{f_s}(u)$ to the basic states of
the representation $D^{f_m}(u)$ (in short, the Clebsch--Gordan
coefficients for the coupling $(f_kf_s)f_m \, )$. In the
fractional--parentage terminology, these SU($l$) Clebsch--Gordan
coefficients represent the $k$--particle coefficients of fractional
parentage of the $m$--particle states, $C_{ f_k v_k f_s v_s }^{ f_m
v_m } = [ k v_k s v_s | \} m v_m ]$. The origin of the combinatorial
factor in Eq.~(\ref{cfp}) and further details are explained in
Appendix~\ref{Anv}.

The pair $A^+(k v_k)A(k v'_k)$ of $k$--particle interaction operators
transforms according to the direct product of the irreducible SU($l$) \
representations $D^{f_k}(u)$ and ${\cal D}^{{\bar f}_k}(u)$. With the
help of an appropriate Clebsch--Gordan transformation, this product can
be reduced to a direct sum
\begin{equation}
D^{f_k}(u) \times {\cal D}^{{\bar f}_k}(u)
= \sum_b D^{g_b}(u)
\label{red}
\end{equation}
of irreducible representations $D^{g_b}(u)$ of Young tableaux $g_b$ .
These are uniquely specified by the ``bodyness'' quantum
number~\cite{mon75} $b$ which assumes the values $b = 0, \ldots , k$.
For Fermions, we have $g_b = (0^{b-1}1$$0^{l-2b-1}10^{b-1})$
while for Bosons, $g_b = (b 0^{l-3} b)$. (In keeping with ABRW, we
confine ourselves for Fermions to the case of less than half filling,
$2m \le l$.) The set of basic $k$--particle interactions $B_k(bw_b)$
which transform according to the irreducible representations
$D^{g_b}(u)$ of SU($l$),
\begin{equation}
T(u) B_k(b w_b) T^+(u)= \sum_{ w'_b }
D_{ w'_b w_b}^{g_b} (u) B_k(b w'_b ) \; ,
\label{TBT}
\end{equation}
is given by
\begin{equation}
B_k(b w_b)
= \sum_{v_k v'_k}
A^+(k v_k) A(k v'_k)
C_{ f_k v_k {\bar f}_k v'_k }^{ g_b w_b } \; .
\label{Bkb}
\end{equation}
The SU($l$) Clebsch--Gordan coefficients $C_{ f_k v_k {\bar f}_k v'_k
}^{ g_b w_b }$ accomplish the reduction. The indices $w_b$ label the
individual states (rows) of the SU($l$) representation $D^{g_b}(u)$ of
Young tableau $g_b$. These running indices range from one to $M_b = N_b^2 -
N_{b-1}^2$, the dimension of the representation $D^{g_b}(u)$. The
representations of the Young tableaux $g_b$ are self--conjugate and
integer. Therefore, we can assume that the $D^{g_b}(u)$ are real and
that the $B_k(b w_b)$ are Hermitean, $B_k^+(b w_b) = B_k(b w_b)$. No
other assumptions about the choice of the irreducible representations
$D^{g_b}(u)$ are made.

Eq.~(\ref{Bkb}) implies the relations
\begin{equation}
\langle \, B_k(b w_b) \rangle_k
= \delta_{b0} \sqrt{N_k} \; ,
\qquad
\langle \, B_k( b w_b ) B_k( b' w'_{b'}) \, \rangle_k
= \delta_{ b b' } \delta_{ w_b w'_{b'} } \; .
\label{Bkb-pro}
\end{equation}
The symbol $\langle O \rangle_t$ denotes the trace of $O$ in the space
of $t$-particle states. According to the Wigner-Eckart theorem for the
group SU($l$), the matrix elements of the operators $B_k( b w_b )$
with respect to the $m$-particle states $| \, m \, v_m \rangle$ are
products of the reduced matrix elements $\langle m || B_k(b) || m
\rangle$ and of the Clebsch-Gordan coefficients $C_{ f_m v_m {\bar f}_m
v'_m }^{ g_b w_b }$ of SU($l$) for the reduction of the direct product
$D^{f_m}(u) \times {\cal D}^{ {\bar f}_m }(u)$ into the direct sum
of the representations $D^{g_b}(u)$,
\begin{equation}
\langle \, m v_m \,| B_k(b w_b) |\, m v'_m \, \rangle
= \langle m || B_k(b) || m  \rangle
C_{ f_m v_m {\bar f}_m v'_m }^{ g_b w_b } \; .
\label{WE}
\end{equation}
The definition~(\ref{WE}) of the reduced matrix elements generalizes
the canonical definition of the reduced matrix elements of the
irreducible tensor operators of SU($2$) given by Racah~\cite{rac42}
to the present case. The dimension--dependent factors $M_b^{-1/2}$ are
absorbed in the reduced elements. For further details on the
Wigner-Eckart theorem for SU($l$), we refer the reader to the book by
Lichtenberg~\cite{lic78} and the references given therein.

The norm $\langle \, B_k^2 (b w _b) \, \rangle_m$ is independent of
$w_b$ and equal to the square of the reduced element. For later use we
note that the norm of any properly normalized $k$--particle operator
$S_k(b)$ of bodyness $b$ has this same value: Making use of
Eq.~(\ref{WE}), we find that every linear form $S_k(b)$ of $B_k(b w_b)$
which satisfies the normalization condition $\langle S_k^+(b) S_k(b)
\rangle_k = 1$, obeys
\begin{equation}
\langle \, S_k^+ (b) S_k(b) \, \rangle_m
= \langle\, m || B_k(b) || m \, \rangle^2 \; .
\label{S-nor}
\end{equation}

The matrices $\langle \, m v_m \,| B_k(b w_b) |\, m v'_m \, \rangle$
play an important role in the sequel. Therefore, it may be useful to
give a physical interpretation of the concept of bodyness, and to
establish the relation between these matrices and the corresponding
quantities appearing in ABRW. As an example, we consider the simplest
version of a pair of creation and annihilation operators: We take $k
= 1$, i.e., we consider the pair $a_i^{\dagger} a_j$. For $i \neq j$,
this operator has bodyness $b = 1$: It moves one particle from the
single--particle state $j$ to a different single--particle state $i$.
For $i = j$, on the other hand, the operator can be written as the sum
of two terms with $b = 0$ and $b = 1$, respectively. Indeed, we can
decompose $a_i^{\dagger} a_i$ into a traceless part $a_i^{\dagger}
a_i - (1/l) \sum_{j = 1}^l a_j^{\dagger} a_j$ and the remainder,
$(1/l) \sum_{j = 1}^l a_j^{\dagger} a_j$. By virtue of
Eq.~(\ref{Bkb-pro}), the traceless part has bodyness $b = 1$, while
the remainder has bodyness $b = 0$. By construction, the interaction
terms of bodyness $b=k$ cannot be simulated by an interaction of lower
particle rank and represent the genuine $k$--body interaction which
generically changes the state of not less than $k$ particles. The
example just considered suggests, and a more detailed consideration
shows, that the matrices  $\langle \, m v_m \,| B_k(b w_b) |\, m v'_m
\, \rangle$ bear a close relationship with the Hermitean matrices
$C^{s a}_{\mu \nu}$ of ABRW. `Indeed, the sets of indices $(b w_b)$
and $(s a)$ can be identified because the dimensions of both sets are
the same and given by $M_b$. The differences between the matrices
$\langle \, m v_m \,| B_k(b w_b) |\, m v'_m \, \rangle$ and the
matrices $C^{s a}_{\mu \nu}$ of ABRW are: (i) We admit any basis and
drop the specialization to Slater determinants or product states
employed by ABRW. (ii) The normalization condition imposed by ABRW on
the matrices $C^{s a}_{\mu \nu}$ differs from our Eq.~(\ref{Bkb-pro}).
(iii) The matrices $\langle \, m v_m \,| B_k(b w_b) |\, m v'_m \,
\rangle$ are introduced group--theoretically, while the matrices
$C^{s a}_{\mu \nu}$ were constructed explicitly. We return to this
comparison in Section~\ref{mom} below.

\section{The Embedded Ensemble}
\label{ens}

After the preliminary steps of Section~\ref{for}, we turn to a
group--theoretical classification of the embedded ensemble EGUE($k$).
This ensemble describes $m$ identical particles distributed over $l$
degenerate single--particle states which interact through a random
$k$--particle interaction of GUE type,
\begin{equation}
W(k) = \sum_{v_k v'_k} W_{v_k v'_k}(k) A^+(k v_k) A(k v'_k) \; .
\label{Wk}
\end{equation}
The coefficients $W_{v_k v'_k}(k)$ are $N_k^2$ independent Gaussian
random variables $W_{v_k v'_k}(k) = (W_{v'_k v_k}(k))^\ast$ with
moments
\begin{equation}
\overline{ W_{v_k v'_k}(k) } = 0 \; , \qquad
\overline{ W_{ v_k v'_k }(k) W_{ {\tilde v}_k {\tilde v}'_k }(k) }
= \frac{ \lambda^2 }{ N_k }
\delta_{v_k {\tilde v}'_k } \delta_{ {\tilde v}_k v'_k } \; .
\label{Wk-mom}
\end{equation}
The overbar denotes the average over the ensemble. Obviously, we must
have $m \geq k$.

The analysis of EGUE($k$) is simplified when we express the interaction
$W(k)$ in terms of the operators $B_k(b \ w_b)$ introduced in
Eq.~(\ref{Bkb}). For brevity we write $\kappa = bw_b$, with $\kappa=0$
referring to the case $b=0$. Making use of the orthonormality of the
traces $\langle \, B_k(\kappa) B_k(\kappa') \, \rangle_k$ (cf.
Eq.~(\ref{Bkb-pro})), we find
\begin{equation}
W(k) = \sum_{\kappa} B_k(\kappa)W_k(\kappa) \; ,
\qquad W_k(\kappa) = \langle \, W(k) B_k(\kappa) \, \rangle_k \; .
\label{Wk-exp}
\end{equation}
This expansion decomposes the interaction $W(k)$ into parts of
well--defined bodyness $b$ with $b = 0, \ldots , k$. For $l$ very
large compared to $m$, the terms of highest bodyness $b = k$ dominate
and yield the main contribution to the average norm $ \overline{ \langle
\, W^2(k) \rangle_m }$.

The matrix representation of the ensemble in the $m$--particle space
has the form
\begin{equation}
H_{v_m v'_m}(k) = \langle \, m v_m |W(k)| m v'_m \, \rangle \; .
\label{Hk}
\end{equation}
The dimension of the matrices $H(k)$ is equal to the dimension
$N=N_m$ of the $m$--particle Hilbert space. We recall that we do not
make any assumption about the basis except that the states $| m v_m
\, \rangle$ are orthonormal. An explicit expression for the matrices
$H(k)$ can be found with the help of the fractional--parentage
expansion Eq.~(\ref{cfp}) of the $m$--particle states. With $s=m-k$,
we find from Eq.~(\ref{Wk}) (for more details we refer to 
Appendix~\ref{Anv}) 
\begin{equation}
H_{v_n v'_n}(k)
= {m \choose k} \sum_{v_k v'_k v_s}
( C_{f_k v_k  f_s v_s}^{ f_m v_m } )^\ast \,
C_{ f_k v'_k  f_s v_s}^{ f_m v'_m } \,
W_{v_k v'_k}(k) \; .
\label{Hk-cfp}
\end{equation}
The dependence of the matrix elements $H_{v_m v'_m}(k)$ of $H(k)$
on the interaction matrix elements $W_{v_k v'_k}(k)$ is, thus,
governed by the SU($l$) Clebsch-Gordan coefficients 
$C_{f_k v_k f_s v_s}^{ f_m v_m }$ for the coupling $(f_k f_s)f_m$.
The expansion Eq.~(\ref{Wk-exp}) introduces
another parametrization of $W(k)$ in terms of the coefficients
$W_k(\kappa)$. The new parameters $W_k(\kappa)$ are real and, by
Eqs.~(\ref{Wk-mom}) and (\ref{Wk-exp}), are independent Gaussian
random variables with moments
\begin{equation}
\overline{ W_k(\kappa) } = 0 \; , \qquad
\overline{ W_k(\kappa) W_k(\kappa') }
= \frac{ \lambda^2 }{N_k} \delta_{\kappa \kappa'} \; .
\label{Wkb-mom}
\end{equation}
In the new parametrization, $H_{v_n v'_n}(k)$ is given by
\begin{equation}
H_{v_m v'_m}(k) = \sum_{\kappa}
\langle \, m v_m | B_k(\kappa) | m v'_m  \, \rangle
W_k(\kappa) \; .
\label{Hk-exp}
\end{equation}

The average of any function $F(H(k))$ of the Hamiltonian $H(k)$
over the ensemble is given by the integral
\begin{equation}
\overline{ F(H(k)) }
= \int \mbox{d} \mu( W(k) ) P( W(k) ) F( H(k) ) \; ,
\label{ave}
\end{equation}
where $\mbox{d}\mu( W(k) )$ denotes the product of differentials of
the matrix elements $W_{v_k v'_k}(k)$, and $P( W(k) )$ the
probability density
\begin{equation}
P( W(k) ) = P_0 \exp \Biggl \{
- \frac{N_k} {2 \lambda^2} \langle \, W^2(k) \, \rangle_k \Biggr \}
\; , \qquad P_0 = \Biggl( \frac{N_k}{ 2 \pi \lambda^2 }
\Biggr)^{ N_k^2 / 2 } \; .
\label{PWk}
\end{equation}
In the parametrization of Eq.~(\ref{Hk-exp}),
$\langle \, W^2(k) \, \rangle_k = \sum_{\kappa} W_k^2(\kappa)$,
and the measure $ \mbox{d}\mu( W(k) ) $ simplifies to the product
of differentials of $W_k(\kappa)$.

The free Gaussian unitary ensemble in $N$ dimensions (the GUE) is
invariant under unitary transformations. More precisely, let $U$
denote an arbitrary unitary matrix of dimension $N$, $U \in
\mbox{U}(N)$. Then, with $H$ a GUE Hamiltonian, ${\hat H} = U H U^+$
is also a member of the ensemble and appears in the ensemble with 
the same weight as the Hamiltonian $H$. EGUE($k$) does not have this
symmetry. However, EGUE($k$) does possess the $U(N_k)$ symmetry of
the Gaussian unitary ensemble for $k$ particles. Indeed, EGUE($k$)
is invariant under the unitary transformation
\begin{equation}
{\hat W}_{v_k v'_k}(k)
= \sum_{ {\tilde v}_k {\tilde v'}_k }
U_{v_k {\tilde v}_k } W_{ {\tilde v}_k {\tilde v}'_k }(k)
(U^+)_{ {\tilde v}'_k v_k } \;
\label{Wk-tran}
\end{equation}
of the interaction matrices $W_{v_k v'_k}(k)$ by any $N_k \times N_k$
unitary matrix $U \in \mbox{U}(N_k)$. By this symmetry transformation,
the Hamiltonian $H(k)$ containing the coeficients $W(k)$ is replaced
by the Hamiltonian ${\hat H}(k)$ containing the coefficients ${\hat
W}(k)$. The ensemble average remains invariant, $\overline{F(H(k))} =
\overline{F({\hat H}(k)) }$. For the special matrices $U=D^{f_k}(u)$
which belong to the SU($l$) representation $D^{f_k}(u)$, the
Hamiltonian ${\hat H}(k)$ takes the simple form ${\hat H}(k) =
D^{f_m}(u) H(k) [D^{f_m}(u)]^+$ and is unitarily equivalent to $H(k)$,
a property which is lacking for a generic $U \in \mbox{U}(N_k)$.
For the parametrization Eq.~(\ref{Hk-exp}), the matrices $H(k)$ are
replaced by
\begin{equation}
{\hat H}_{v_m v'_m}(k) = \sum_{\kappa \kappa'}
\langle \, m v_m | B_k(\kappa') | m v'_m  \, \rangle
W_k(\kappa) \Delta_{\kappa' \kappa}^k(U) \; .
\label{Hk-tran-exp}
\end{equation}
Here $\Delta^k(U)$ denotes the matrix representation of 
$U \in \mbox{U}(N_k)$, carried by the matrices 
$\langle \, k v_k |B_k(\kappa)|k v'_k \, \rangle$, 
\begin{equation}
\sum_{ {\tilde v}_k {\tilde v}'_k} U_{v_k {\tilde v}_k}
\langle \, k {\tilde v}_k |B_k(\kappa)|k {\tilde v}'_k \, \rangle
(U^+)_{ {\tilde v}'_k v'_k}
= \sum_{\kappa'}\Delta_{\kappa' \kappa}^k(U)
\langle \, k v_k |B_k(\kappa')|k v'_k \, \rangle \; .
\label{DkU-def}
\end{equation}
This representation is unitary and real. It is the direct sum of two
irreducible representations, the identity representation carried by
the matrix of $B_k(0)$, and the $(N_k^2 -1)$--dimensional irreducible
representation of the Young tableau $(1 0^{N_k^2-3} 1)$ carried by the
matrices of $B_k(\kappa)$ with $b>0$. For $U = D^{f_k}(u)$, the
matrices $\Delta^k(U)$ simplify to $\Delta_{\kappa' \kappa}^k(U) =
\delta_{bb'}D_{w'_b w_b}^{g_b}(u)$. More details on the matrices
$\Delta^k(U)$ are given in Appendix~\ref{DkU-pro}. The invariance of
the ensemble under U($N_k$) implies that whenever the integrals over
$W_k(\kappa)$ are performed and the ensemble averages are expressed
in terms of the matrix elements of the operators $B_k(\kappa)$, the
resulting formulae must be invariant under the replacement of
$B_k(\kappa)$ by
\begin{equation}
{\hat B}_k(\kappa)
= \sum_{\kappa'} \Delta_{\kappa' \kappa}^k(U) B_k(\kappa') \; ,
\qquad U \in \mbox{U}(N_k) \; .
\label{Bkb-tran-A}
\end{equation}

As pointed out in ABRW, the properties of EGUE($k$) are closely related
to those of the ``dual'' ensemble EGUE($s$)=EGUE($m-k$). The dual
ensemble describes $m$ particles interacting by an $s$-particle GUE
interaction $W(s)$. The Clebsch--Gordan coefficients $C_{ f_k v_k f_s
v_s }^{ f_m v_m }$ and $C_{ f_s v_s f_k v_k }^{ f_m v_m }$ differ at
most by a phase independent of the $v$'s. Therefore, the Hamiltonian
$H(s)$ of the dual ensemble can be written as
\begin{equation}
H_{v_m v'_m}(s)
= {m \choose s} \sum_{v_k v_s v'_s}
( C_{ f_k v_k f_s v_s }^{ f_m v_m } )^\ast \,
C_{ f_k v_k f_s v'_s }^{ f_m v'_m } \,
W_{v_s v'_s}(s) \; , \quad s = m-k \; .
\label{Hs-cfp}
\end{equation}
We use the link established by duality between EGUE($k$) and EGUE($s$)
in Section~\ref{mom}.

\section{The Matrix of Second Moments}
\label{mom}

In ABRW, the matrix of second moments plays a central role for
the analysis of EGUE($k$). In the present Section, we address this
matrix from the point of view of group theory.

The matrix elements $H_{v_m v'_m}(k)$ are Gaussian random variables.
Therefore, all information necessary for the evaluation of averages
over the ensemble EGUE($k$) is contained in the matrix of second
moments,
\begin{equation}
\label{secm}
A_{v_m {\tilde v}'_m, {\tilde v}_m v'_m}(k) = \overline{ H_{v_m
v'_m}(k) H_{ {\tilde v}_m {\tilde v}'_m }(k) } \; .
\end{equation} 
As shown by ABRW, central information on the properties of EGUE($k$)
can be deduced if a ``generalized eigenvalue expansion'' for this
matrix can be found. In particular, the application of Efetov's
supersymmetric averaging technique \cite{efe83} to EGUE($k$) becomes
possible. Such an expansion has the form
\begin{equation}
\label{eigv}
A_{v_m {\tilde v}'_m, {\tilde v}_m v'_m}(k) = \sum_{\alpha} C_{v_m
{\tilde v}'_m}^{\alpha} \Lambda^{\alpha} C_{ {\tilde v}_m v'_m}^{\alpha} \; .
\end{equation}
Here, the ``eigenvalues'' $\Lambda^{\alpha}$ must be positive, and the
``eigenvectors'' $C_{v_m v'_m}^{\alpha}$ should be Hermitean. In ABRW,
the expansion~(\ref{eigv}) was constructed explicitly.

We show now that this expansion is closely related to the bodyness
expansion of the Hamiltonian $H(s)$ of the dual ensemble EGUE($s$).
Following ABRW, we note first that the second moments of the
Hamiltonian matrix elements of the ensembles EGUE($k$) and EGUE($s$)
satisfy the ``duality'' relation
\begin{equation}
\overline{ N_k H_{v_m v'_m}(k) H_{ {\tilde v}_m {\tilde v}'_m }(k) }
= \overline{ N_s H_{v_m {\tilde v}'_m}(s) H_{ {\tilde v}_m v'_m }(s)}
\; .
\label{mom-dua}
\end{equation}
Here, the average on the left--hand side is over EGUE($k$), the
average on the right--hand side is over EGUE($s$). The duality
relation can be verified by expressing the matrix elements of $H(k)$
and $H(s)$ in terms of the coefficients of fractional parentage
$C_{f_k v_k f_s v_s}^{f_m v_m}$ (Eqs.~(\ref{Hk-cfp}) and
(\ref{Hs-cfp})), and taking the ensemble averages of the products of
the interaction matrix elements $W_{v_k v'_k}(k)$ and
$W_{v_s v'_s}(s)$. The duality relation is robust with respect to a
truncation of Hilbert space: It remains valid when some of the basis
states $|m v_m \, \rangle$ are excluded. However, this relation is
obviously violated when the random $k$--particle interactions are
modified  in such a way that their ensembles lose their unitary
symmetry so that Eq.~(\ref{Wk-mom}) is not valid any more. An
important example of such  a modification is the restriction of the
interaction to the terms of the highest possible
bodyness~\cite{pat00}. 

Another important observation relates to duality. The duality
relation allows us to evaluate the second moments of EGUE($k$) in
terms of the  second moments of EGUE($s$). This yields
the moments $\overline{ H_{v_m v'_m}(k) H_{ {\tilde v}_m {\tilde
v}'_m }(k) }$ as bilinear forms of matrix elements of operators
$B_s(\kappa)$. From the U($N_s$) symmetry of EGUE($s$), the
resulting expressions are invariant under the replacement of
$B_s(\kappa)$ by
\begin{equation}
{\hat B}_s(\kappa) 
= \sum_{\kappa'}\Delta_{\kappa'\kappa}^s(U) B_s(\kappa') \; 
\label{hBs-tra}
\end{equation}
for any $U \in \mbox{U}(N_s)$. Here, $\Delta^s(U)$ denotes the 
transformation matrix defined analogously to Eq.~(\ref{DkU-def}). 
The same ``dual'' U($N_s$) invariance holds obviously for any 
ensemble average $\overline{F(H(k))}$ calculated from the second
moments $\overline{ H_{v_m v'_m}(k) H_{ {\tilde v}_m {\tilde v}'_m
}(k) }$ evaluated in this way. This concerns, in particular, the
supersymmetric $n$--point generating functions of EGUE($k$).

We turn to the matrix of second moments $\overline{ H_{v_m
{\tilde v}'_m}(s) H_{ {\tilde v}_m v'_m }(s)}$ of the Hamiltonian
$H(s)$. We decompose the interaction $W(s)$ into a sum of terms with
well--defined bodyness,
\begin{equation}
H_{v_m v'_m}(s) = \sum_{\kappa}
\langle \, m v_m | B_s(\kappa) | m v'_m  \, \rangle
W_s(\kappa) \; .
\label{Hs-exp}
\end{equation}
We take the average of pairs of the random variables $W_s(\kappa)$ and
express the matrix elements $\langle \, m v_m \,| B_s(\kappa) |\, m
v'_m \, \rangle$ with the help of the SU($l$) Wigner-Eckart formula
Eq.~(\ref{WE}). We use the duality relation Eq.~(\ref{mom-dua}) and
get
\begin{equation}
A_{v_m {\tilde v}'_m, {\tilde v}_m v'_m}(k)
= \overline{ H_{v_m v'_m}(k) H_{ {\tilde v}_m {\tilde v}'_m } (k) }
= \sum_{b w_b}^s C_{v_m {\tilde v}'_m}^{b w_b}
\Lambda^b(k) C_{ {\tilde v}_m v'_m}^{b w_b}
\label{mom-exp}
\end{equation}
where
\begin{equation}
\Lambda^b(k) =
\frac{\lambda^2}{N_k}
\langle \, m || B_s(b) || m \, \rangle ^2 \; ,
\qquad C_{v_m v'_m}^{b w_b}
= C_{ f_m v_m {\bar f}_m v'_m}^{g_b w_b} \; .
\label{Labk-Cbv}
\end{equation}
The upper limit at the summation symbol indicates that the summation
over $b$ is restricted to $b \le s$. Equation~(\ref{Labk-Cbv})
constitutes the eigenvalue expansion of the matrix $ A_{v_m {\tilde
v}'_m, {\tilde v}_m v'_m}(k)$, with $\Lambda^b(k) \geq 0$ the
eigenvalues and $C^\kappa = ( C_{v_m v'_m}^\kappa )$ the Hermitean
eigenvectors. The eigenvalues $\Lambda^b(k)$ are labelled only by
the bodyness quantum number $b$.

We demonstrate the dual U($N_s$) invariance of this spectral decomposition
of $A_{v_m {\tilde v}'_m, {\tilde v}_m v'_m}(k)$ by writing
Eq.~(\ref{mom-exp}) as (in the sequel we suppress the upper limit $s$
at the summation symbol)
\begin{equation}
A_{v_m {\tilde v}'_m, {\tilde v}_m v'_m}(k)
= \frac{1}{N} \sum_\kappa V_{v_m {\tilde v}'_m}^\kappa(k)
V_{ {\tilde v}_m v'_m}^\kappa(k) \; .
\label{A-VV}
\end{equation}
Here $V^\kappa(k)$ stands for 
\begin{equation}
V_{v_m v'_m}^{\kappa}(k) 
= \sqrt{N \Lambda^\kappa(k) } C_{v_m v'_m}^\kappa 
= \lambda \sqrt{N/N_k} 
\langle \, m v_m|B_s(\kappa)| m v'_m \, \rangle \; .
\label{Vka}
\end{equation}
Under the U($N_s$) symmetry transformations Eq.~(\ref{hBs-tra}), 
the matrices $V^\kappa(k)$ transform as 
\begin{equation}
{\hat V}^\kappa(k) 
= \sum_{\kappa'}\Delta_{\kappa' \kappa}^s(U)V^{\kappa'}(k) \; ,
\qquad U \in \mbox{U}(N_s) \; .
\label{Vb-tra}
\end{equation}
Since the matrix $\Delta^s(U)$ is unitary and real, the sum over
$\kappa$ appearing in Eq.~(\ref{A-VV}) remains invariant under this
transformation. This fact proves the invariance of $A_{v_m {\tilde
v}'_m, {\tilde v}_m v'_m}(k)$.

The expansion~(\ref{mom-exp}) gives the matrix of the second moments
in terms of the SU($l$) reduced matrix elements $ \langle \, m ||
B_s(b) || m \, \rangle$ and in terms of the SU($l$) Clebsch-Gordan
coefficients $C_{ f_m v_m {\bar f}_m v'_m}^{g_b w_b}$. The reduced
matrix elements of $B_s(\kappa)$ yield the eigenvalues and can be
expressed in terms of the SU($l$) recoupling coefficients $\langle \,
( (f_m {\bar f}_k ) f_s f_k )f_m | ( f_m (f_k {\bar f}_k) g_b) f_m \,
\rangle$. However, as indicated by Eq.~(\ref{S-nor}), these matrix
elements can be evaluated even more directly by constructing suitable
interactions of the required bodyness and particle rank, and by
calculating the corresponding trace. We present this calculation in
Appendix~\ref{red-ele}. In the Fermionic case this yields
\begin{equation}
\Lambda^b(k) = \frac{\lambda^2}{N_k}
{ {m - b} \choose k }{ {l - m + k - b} \choose k } \; ,
\label{Labk-fer}
\end{equation}
whereas in the Bosonic case we get
\begin{equation}
\Lambda^b(k) = \frac{\lambda^2}{N_k}
{ {m - b} \choose k }{ {l + m + b - 1} \choose k } \; .
\label{Labk-bos}
\end{equation}
According to Eq.~(\ref{mom-dua}), the eigenvalue expansions of the
matrices of second moments of the two dual ensembles EGUE($k$) and
EGUE($s$) are related by
\begin{equation}
N_k \sum_{\kappa}
C_{v_m {\tilde v}'_m}^\kappa \Lambda^b(k) C_{{\tilde v}_m v'_m}^\kappa
= N_s \sum_{\kappa} 
C_{v_m v'_m}^\kappa \Lambda^b(s) C_{ {\tilde v}_m {\tilde v}'_m}^\kappa \; .
\label{exp-dua}
\end{equation}

When we compare these results with the eigenvalue expansion obtained
in ABRW we see that, aside from normalization factors, the two
expansions agree. Indeed, the eigenvalues found in ABRW agree
with Eqs.~(\ref{Labk-fer}) and~(\ref{Labk-bos}), 
except for the factors $\lambda^2/N_k$ which are missing in ABRW.
This difference is caused by the fact that in ABRW,
the second moment of the interaction matrix elements is normalized
to unity, whereas in Eq.~(\ref{Wk-mom}) we have used  the standard
GUE normalization for $k$ interacting particles. We conclude that
the eigenvalues of ABRW are identical to the squares of the
reduced matrix elements $\langle m || B_s(b) || m \rangle$, and that
the eigenvalues $C^{s a}_{\mu \nu}$ are, aside from a factor
$\sqrt{N}$ due to the difference in normalization, given by the
Clebsch--Gordan coefficients $C^{bw_b}_{v_m v'_m}$. The factors
$N_k$ and $N_s$ appearing on either side of the duality relations
Eqs.~(\ref{mom-dua}) and (\ref{exp-dua}) are likewise absent in
ABRW. Again, this is due to the difference in normalization of
the random variables.

\section{Invariance under the Group U($N$)}
\label{inv}

We have stressed in Section~\ref{ens} that for the free GUE ensemble
in $N$ dimensions (or, equivalently, for EGUE($m$)), the matrix of
second moments is invariant under the transformation of the
Hamiltonian by any unitary $N \times N$ matrix $U \in \mbox{U}(N)$, while
for $k < m$, the matrix of the second moments of EGUE($k$) does not
have this symmetry property. In the present section, we display the
broken U($N$) symmetry of EGUE($k$) explicitly.

To determine the U($N$)--invariant part $A^{(0)}(k)$ of the matrix
of second moments $A(k)$, we consider a transformation $U \in
\mbox{U}(N)$ with $H(k) \rightarrow {\hat H}(k) = UH(k)U^+$. Under
this transformation, the components  $C_{v_m v'_m}^\kappa$ of the
eigenvector $C^\kappa$ transform like the components of an U($N$) tensor
with one covariant index and one contravariant index. Starting from
Eq.~(\ref{mom-exp}), separating the U($N$)--invariant part of $ C_{v_m
{\tilde v}'_m}^\kappa C_{ {\tilde v}_m v'_m }^\kappa $, and working out
the sum over $\kappa$ with the help of the duality relation
Eq.~(\ref{exp-dua}), we find that the U($N$)--invariant part
$A^{(0)}(k)$ has the form
\begin{eqnarray}&&
A_{ v_m {\tilde v}'_m, {\tilde v}_m v'_m }^{(0)}(k)
= \frac{N \Lambda^0(k)}{N^2 - 1}
\Biggl\{
\delta_{ v_m {\tilde v}'_m } \delta_{ {\tilde v}_m v'_m }
\Biggl(1 - S(k) \Biggr)
\qquad \qquad \qquad \qquad
\nonumber \\ && \qquad \qquad \qquad \qquad \qquad \qquad
+ \delta_{ v_m v'_m } \delta_{ {\tilde v}_m {\tilde v}'_m }
N \Biggl( S(k) - \frac{1}{N^2} \Biggr) \Biggr\} \; ,
\label{A0k}
\end{eqnarray}
where
\begin{equation}
\Lambda^0(k) = \frac{1}{N} \, \overline{ \langle H^2(k) \rangle_m } \; ,
\qquad
S(k) = \frac{1}{N}\frac{N_s \Lambda^0(s)}{N_k \Lambda^0(k)}
= \frac{1}{N} \frac{ \overline{  \langle \, H(k)  \, \rangle_m^2 \; } }
{ \overline{ \langle \, H^2(k)  \, \rangle_m } } \; .
\label{La0k-Pk}
\end{equation}
The U($N$)--invariant part of $A(k)$ is thus specified by the square
$\Lambda^0(k)$ of the average spectral width and by the ratio $S(k)$.
This ratio was introduced in ABRW as a measure of the ergodicity of
the spectral centroids. We remark in parentheses that the converse is
also true: The average spectral width $\sqrt{\Lambda^0(k)}$ and the
ratio $S(k)$ are governed entirely by $A^{(0)}(k)$. There are no
contributions either to $\Lambda^0(k)$ or to $S(k)$ arising from the
U($N$)--non--invariant part of $A(k)$.

The result displayed in Eq.~(\ref{A0k}) is somewhat unexpected. Indeed,
the U($N$)--invariant matrix $A^{(0)}(k)$ is the sum of two terms,
$A^{(0)}(k) = A^{\mbox{\scriptsize{G}}}(k) + A^{\mbox{\scriptsize{D}}}(k)$. 
The first,
\begin{equation}
A_{v_m {\tilde v}'_m, {\tilde v}_m v'_m }^{\mbox{\scriptsize{G}}}(k)
= \frac{1}{N} \Lambda^{\mbox{\scriptsize{G}}}(k)
\delta_{ v_m {\tilde v}'_m } \delta_{ {\tilde v}_m v'_m }
\label{AGk}
\end{equation}
has the expected form: The Kronecker symbols carry the same indices
as in the matrix of second moments of the free Gaussian unitary
ensemble of $N \times N$ matrices or, equivalently, in EGUE($m$). The
average spectral width $\Lambda^{\mbox{\scriptsize{G}}}$ corresponding to the
GUE described by this term 
is given by
\begin{equation}
\Lambda^{\mbox{\scriptsize{G}}}(k) = \Lambda^0(k) N^2 
\frac{1 - S(k)}{N^2 - 1} \; .
\label{LaGk}
\end{equation}
The form of the second term,
\begin{equation}
A^{\mbox{\scriptsize{D}}}_{ v_m {\tilde v}'_m, {\tilde v}_m v'_m }(k) 
= \Lambda^{\mbox{\scriptsize{D}}}(k) \,
\delta_{ v_m v'_m } \delta_{ {\tilde v}_m {\tilde v}'_m } \; ,
\label{ADk}
\end{equation}
is unexpected. Indeed, the indices on the Kronecker deltas occur
in the same way as in the matrix of second moments of the degenerate
Gaussian ensemble of multiples of the $N \times N$ unit matrix or,
equivalently, in EGUE($0$). The average spectral width
corresponding to the degenerate Gaussian ensemble described by  
this term is given by
\begin{equation}
\Lambda^{\mbox{\scriptsize{D}}}(k) = \Lambda^0(k) 
\frac{N^2 S(k) - 1}{N^2 - 1} \; .
\label{LaDk}
\end{equation}
The ensemble EGUE($k$) thus differs from the free GUE ensemble not
only by the presence of the U($N$)--non--invariant part but also in
the structure of the U($N$)--invariant part $A^{(0)}(k)$. This part
contains, in addition to the expected term 
$A^{\mbox{\scriptsize{G}}}(k)$, also the term
$A^{\mbox{\scriptsize{D}}}(k)$. The relative importance of the terms 
$A^{\mbox{\scriptsize{G}}}(k)$ and $A^{\mbox{\scriptsize{D}}}(k)$
is governed by the parameter $S(k)$. For fixed $m$ and $l$, this
parameter decreases with increasing $k$ from the value unity at $k=0$
to the value $N^{-2}$ at $k=m$, with $S = N^{-1}$ at $k=m/2$. At
$k=0$, the GUE term $A^{\mbox{\scriptsize{G}}}(k)$ is absent since 
$\Lambda^{\mbox{\scriptsize{G}}}(0) = 0$. With
increasing $k$, the width $\Lambda^{\mbox{\scriptsize{G}}}$ 
increases, and for $k \ge m/2$ and large $N$, becomes dominant, 
$\Lambda^{\mbox{\scriptsize{G}}}(k) = \Lambda^0(k)$ up
to terms of order $\mbox{O}(N^{-1})$. For finite $N \gg 1$ and small $k$,
the U($N$)--invariant part $A^{(0)}(k)$ favors a Gaussian spectral
shape and non--ergodic behavior, while for $k > m/2$, the favored spectral
shape becomes semicircular and the spectral fluctuations, ergodic.

The U($N$) symmetry of EGUE($k$) is obviously violated most strongly
when $k$ is close to $m/2$. A convenient measure for the degree of
symmetry breaking is given by the ratio $P(k)$ of the norm of the
U($N$)--invariant part $A^{(0)}(k)$ and of the norm of $A(k)$. From 
Eqs.~(\ref{A0k}) and (\ref{mom-exp}) we find that $P(k)$ is given by
\begin{equation}
P(k)
= \frac{ \mbox{Tr}( [A^{(0)}(k)]^+ A^{(0)}(k) )}
{ \mbox{Tr}( A^+(k)A(k) ) }
= \frac{ 1 - 2 S(k) + N^2 S^2(k) }{ (N^2 - 1 ) (R(k)/2) } \; .
\label{xik}
\end{equation}
Here $R(k)$ denotes the coefficient
\begin{equation}
R(k)
= \frac{ 2 \sum_b^s M_b ( \Lambda^b(k) )^2 }
{ ( N \Lambda^0(k) )^2 } \;
= \frac{ \overline{ \; \langle \, H^2(k) \, \rangle_m^2 \;} }
{ \; (\overline{ \, \langle \, H^2(k) \, \rangle_m } \;)^2 }
- 1  
\; 
\label{Rk}
\end{equation} 
first introduced in ABRW as the measure of ergodicity of the spectral
widths. The ratio $P(k)$ is symmetric about the point $k=m/2$ and
attains its minimum there.

The results of the present Section cast new light on the most
surprising result of ABRW. There it was found that for $k \ll m$,
the spectral fluctuations of EGUE($k$) differ markedly from
Wigner--Dyson form and tend towards Poissonian behavior. While the
present analysis does not yield direct information on this question,
the fact that for $k \ll m$ the embedded ensemble EGUE($k$) is 
dominated by the U($N$)--invariant term $A^{\mbox{\scriptsize{D}}}(k)$, 
lends additional
plausibility to this result.

\section{Symmetry Properties of the \\ Supersymmetric Generating
Functions}
\label{gen}

We show how symmetry arguments are involved in the use of the
supersymmetry aproach introduced by Efetov \cite{efe83} and
developed by Verbaarschot, Weidenm\"uller and Zirnbauer\cite{ver85}.
We use the same conventions as in Ref.~\cite{ver85}. We address first
the one--point function and its saddle--point solution and then turn
briefly to the generating functions of higher order.

The ensemble average of the one--point function $Z(E,H)$ of energy
$E$ is given by the graded integral
\begin{equation}
\overline{ Z( E, H ) }
= \int \! \mbox{d}\mu(\psi)
\exp \Biggl \{
i \psi^+ \left( E_+ 1 \times 1 + J \right) \psi
- \frac{1}{2} \overline{ \left( \psi^+ (1 \times H) \psi \right)^2 }
\Biggr \} \;
\label{ZEH-psi}
\end{equation}
over the field  $\psi$ of $2N$ components $\psi_{\alpha v_m}$.
The components with $\alpha = 0$ are ordinary complex variables, the
components with $\alpha = 1$ anticommute. The matrices which appear
in the exponent are written as direct products of the $2 \times 2$
matrices acting on the indices $\alpha$ and the $N \times N$ matrices
acting on the indices $v_m$. The complex energy $E_+ = E + i\eta$
contains the infinitesimal term $\eta = 0_+$ introduced to assure
convergence, and $J$ is the source matrix $J_{\alpha v_m,\alpha' v'_m}
= \delta_{\alpha \alpha'} (-)^{\alpha + 1}\, \xi_{v_m v'_m}$. 
For notational simplicity,
we suppress the index $k$ throughout and write $H(k) = H$.

All information about the ensemble is contained in the matrix of second
moments $A$. We first consider a toy model. We omit the
U($N$)--non--invariant part of the matrix $A$. Then, $A$ is given by
the U($N$)--invariant part $A^{(0)} = A^{\mbox{\scriptsize{G}}} 
+ A^{\mbox{\scriptsize{D}}}$ defined in
Section~\ref{inv}. Making use of the transformation
\begin{equation}
\exp \Biggl \{ - \frac{1}{2} \Lambda^{\mbox{\scriptsize{D}}} 
(\psi^+ \psi)^2  \Biggr \}
= \Biggl( \frac{1}{ 2\pi \Lambda^{\mbox{\scriptsize{D}}} } 
\Biggr)^{1/2} \! \int \! \mbox{d} t
\exp \Biggl \{ - \frac{1}{2 \Lambda^{\mbox{\scriptsize{D}}} }
( t^2  +  2it \Lambda^{\mbox{\scriptsize{D}}} \psi^+ \psi )
\Biggr\} \; ,
\label{HST}
\end{equation}
we find that in this case, the function $\overline{Z(E,H)}$ simplifies
to the convolution
\begin{equation}
\overline{ Z^{(0)}(E) }
= \Biggl( \frac{1}{ 2\pi \Lambda^{\mbox{\scriptsize{D}}} } 
\Biggr)^{1/2} \! \int \! \mbox{d}t
\exp \Bigl( - \frac{1}{2 \Lambda^{\mbox{\scriptsize{D}}} } \, t^2
\Bigr) \, \overline{ Z^{\mbox{\scriptsize{G}}}(E-t, 
\Lambda^{\mbox{\scriptsize{G}}} ) } \;
\label{Z0E}
\end{equation}
of the Gaussian of width $\sqrt{\Lambda^{\mbox{\scriptsize{D}}}} $ with
the one--point function $\overline{ Z^{\mbox{\scriptsize{G}}}(E, 
\Lambda^{\mbox{\scriptsize{G}}} ) }$ of the free GUE in $N$ 
dimensions with the average spectral width
$\sqrt{\Lambda^{\mbox{\scriptsize{G}}}} $. The average level density
$\overline{ \rho^{(0)}(E) }$ corresponding to $\overline{ Z^{(0)}(E)}$
has the width $\sqrt{\Lambda^0}$ with
$\Lambda^0 = \Lambda^{\mbox{\scriptsize{G}}} +
\Lambda^{\mbox{\scriptsize{D}}}$. As for the kurtosis of the average
level density, we follow the convention of ABRW: Their quantity
$Q^{(0)}$ equals unity in case the spectral shape is Gaussian, and zero
if the spectrum has semicircular shape. We find $Q^{(0)} = 1 -
( \Lambda^{\mbox{\scriptsize{G}}} / \Lambda^0 )^2 $. Thus, $Q^{(0)}$
decreases from the value one at $k = 0$ to zero at $k = m$. For large
$N$, the average level density is given by the convolution of a
Gaussian of width $\sqrt{\Lambda^{\mbox{\scriptsize{D}}}} $ with the
semicircular distribution of width
$\sqrt{\Lambda^{\mbox{\scriptsize{G}}}}$. 
In general, the actual behavior of $\overline{Z(E,H)}$ is much more 
complex than that of this toy model. However, the toy model may perhaps 
be not far from the truth for $k \ll m$ and for $k$ close to $m$. 
In both cases the non--invariant part of $A$ is comparatively small.

We now address the one--point function in its full generality. With the
help of the eigenvalue expansion of the matrix of second moments $A$ and
the Hubbard--Stratonovich transformation, the quartic term appearing in
the exponent of Eq.~(\ref{ZEH-psi}) can be simplified and the integrals
over $\psi$ can be performed. This yields
\begin{eqnarray} &&
\overline{ Z(E,H) }
= \int \! \mbox{d}\mu(\sigma) \exp \left \{ - \frac{1}{2}N
\sum_\kappa \left \langle (\sigma^\kappa)^2 \right
\rangle \qquad \qquad \right.
\nonumber \\ && \qquad \qquad \left.
- \left \langle  \ \ln \left( E_+ 1 \times 1 +  J 
- \sum_\kappa \sigma^\kappa \times V^\kappa \right)
\right \rangle \right \} \; ,
\label{ZEH-sig}
\end{eqnarray}
where $\mbox{d}\mu(\sigma)$ denotes the measure for integration over
the graded $2 \times 2$ matrices $\sigma^\kappa$ introduced in the
Hubbard-Stratonovich transformation. We use $V^\kappa = \sqrt{N \Lambda^b}
C^\kappa$. The angular brackets denote graded traces. Under a dual unitary
transformation $U \in \mbox{U}(N_s)$, the matrices $V^\kappa$ are 
replaced by the matrices ${\hat V}^\kappa$ introduced
in Eq.~(\ref{Vb-tra}). Since $\Delta^s(U)$ are real unitary matrices, the
transformation coefficients can be absorbed in the new integration
variables ${\hat \sigma}^{\kappa} = \sum_{\kappa'} \Delta_{\kappa 
\kappa'}^s(U) \sigma^{\kappa'} $, with $\sum_{\kappa} 
( {\hat \sigma}^\kappa )^2 
= \sum_{\kappa}( \sigma^\kappa )^2 $. The integral thus remains 
invariant as required.

The same symmetry consideration limits the form of the saddle--point
solution. Indeed, with $\sigma_{\mbox{\scriptsize{sp}}}^\kappa$ 
the saddle--point solution, the dual U($N_s$) invariance implies that 
the saddle--point approximation to the generating function 
\begin{eqnarray} &&
\overline{ Z_{\mbox{\scriptsize{sp}}}(E,H) }
= \exp \left \{
- \left \langle  \ \ln \left( E_+ 1 \times 1 + J 
- \sum_\kappa \sigma_{\mbox{\scriptsize{sp}}}^\kappa \times V^\kappa \right)
\right \rangle \right \} \;
\label{ZspEH-sig}
\end{eqnarray}
must be invariant under the replacement of $V^\kappa$ by ${\hat V}^\kappa$,
Eq.~(\ref{Vb-tra}), for any $U \in \mbox{U}(N_s)$. This condition 
can be fulfilled only when
$\sigma_{\mbox{\scriptsize{sp}}}^\kappa = 0$ for all $b > 0$.
The saddle--point solution found in ABRW has just this form, with
\begin{equation}
\sigma_{\mbox{\scriptsize{sp}}}^\kappa = \delta_{b 0} \, \sigma_0 \; , \qquad
\sigma_0 = \left( \frac{E}{ 2 \sqrt{\Lambda^0 } }
-i \sqrt{ \, 1 - \left( \frac{E}{2 \sqrt{\Lambda^0} }
\right)^2 } \,
\right ) \cdot 1 \; .
\label{SPE-sol}
\end{equation}
We stress that this saddle--point solution exists only by virtue 
of the fact that the basis states $|m v_m \,\rangle$ represent 
a complete set of states of the irreducible representation $D^{f_m}(u)$. 
This can be seen in the following way. As discussed in ABRW, the saddle
--point equation has the form
\begin{equation}
N (\sigma^\kappa_{\mbox{\scriptsize{sp}}})_{\alpha \alpha'} =
\sum_{v_m v'_m}
\left( E \,1\times 1
- \sum_{\kappa'} \sigma_{\mbox{\scriptsize{sp}}}^{\kappa'} 
\times V^{\kappa'} \right)^{-1}
_{\alpha v_m, \alpha' v'_m} V_{v'_m v_m}^\kappa \; .
\label{Z1-speA}
\end{equation}
Substitution shows that $\sigma_{\mbox{\scriptsize{sp}}}^\kappa 
= \delta_{b0} \sigma_0$
is a solution only if the traces of the matrices $V^\kappa$ vanish for
all $b > 0$. These traces differ only by a factor from the traces
$\langle B_m(\kappa) \rangle_m$. Moreover, only scalar quantities can
have a nonzero trace in the complete representation space of an
irreducible representation. Hence this condition is met. However, the
exclusion of some of the states $|m v_m \, \rangle$ from the Hilbert
space of the ensemble would make this argument invalid. The
supersymmetry approach could then still be applied, but a saddle point
would not exist generically.

For the generating functions of higher order, the symmetry properties
of the ensemble manifest themselves in the same way. As shown in detail 
in ABRW, the two--point function can be written as the graded 
integral
\begin{eqnarray} &&
\overline{ Z(E + \omega, E - \omega, H) }
= \int \! \mbox{d} \mu(\sigma)
\exp \left \{  -\frac{1}{2} N \sum_\kappa \langle (\sigma^\kappa)^2
\rangle \qquad \qquad \quad \right.
\nonumber \\ && \left. \qquad
- \left \langle
\ln \left ( ( E \cdot 1 + \omega_{+} L ) \times 1 + J 
- \sum_\kappa \sigma^\kappa \times V^\kappa \right )
\right \rangle
\right \} \;
\label{ZEEH-Tsig}
\end{eqnarray}
over the $4 \times 4$ graded matrices $ \sigma^\kappa = ( \sigma_{\alpha p,
\alpha' p'}^\kappa )$ . The doubling of dimensions of the graded matrices 
reflects the fact that we are dealing with two propagators, the retarded 
propagator of energy $E+\omega$ $(p=1)$, and the advanced propagator of 
energy $E-\omega$ $(p=2)$. The diagonal matrix $L$ with
$L_{\alpha p, \alpha' p'} = \delta_{\alpha \alpha'} \delta_{pp'}
(-)^{p+1}$ distinguishes between the two cases. The matrix $J$
contains the source parameters. As discussed thoroughly in ABRW, the 
integral is dominated by the saddle--point manifold 
\begin{equation}
\sigma_{\mbox{\scriptsize{sp}}}^\kappa = \delta_{b0} T^{-1} \sigma_0 T \; ,
\qquad
\sigma_0 = \frac{E}{ 2 \sqrt{\Lambda^0} } \cdot 1
-i \sqrt{ \, 1 - \left( \frac{E}{2 \sqrt{\Lambda^0}} \right)^2 \,
} \cdot L \; \; .
\label{SP-man}
\end{equation}
Here, $T=(T_{\alpha p, \alpha' p'})$  denote the $4 \times 4$ graded 
matrices belonging to the coset space $\mbox{U}(1,1/2) /$ 
$ [\mbox{U}(1/1) \times \mbox{U}(1/1)]$. The integral as well as the 
saddle--point approximation are manifestly invariant 
with respect to the dual symmetry transformations $U \in \mbox{U}(N_s)$,
Eq.~(\ref{Vb-tra}). 
Saddle--point manifolds of the same structure dominate also the graded 
integrals which represent the generating functions of higher order.

\section{Discussion and Conclusions}
\label{con}

In this paper, we have shown that some of the central results of
ABRW can be deduced with the help of symmetry considerations.
This fact explains why analytical progress in understanding the
properties of EGUE($k$) has been possible at all. In addition, our
work offers a deeper insight into the structure of EGUE($k$). We
mention, in particular, the decomposition in Section~\ref{inv} of
the matrix of second moments into a part which is invariant under
$U \in$ U($N)$ and another which is not.

We mention the following generalizations of our work.

(i) The theory developed in Sections \ref{ens} - \ref{gen} for
EGUE($k$) is based on the SU($l$) expansion formula
Eq.~(\ref{Hk-cfp}). We emphasize that the approach developed in
ABRW and in the present paper is not restricted to this case
but is much more general. Indeed, it applies likewise to ensembles
defined by the same formula but with the SU($l$) Clebsch--Gordan
coefficients replaced by the Clebsch--Gordan coefficients of
another symmetry group G. The will now discuss this case in order
to illuminate the algebraic structure of the approach used in the
present paper.

We introduce a generalized ensemble by considering two independent
systems labelled $j=k$ and $j=s$ whose basic states
$|f_j v_j \rangle$ transform according to the irreducible
representations $D^{f_j}(g)$ of dimension $N_j$ of the group $G$.
We assume for definiteness that the group $G$, $g \in G$, is a compact
simple Lie group; the irreducible representations of $G$ will be labelled
by their highest weights $f$, their basis states by the running indices 
$v$. We assume that a non--trivial interaction $W(k)$ of
GUE type occurs only in system $k$, with $W_{v_k v'_k}(k) = \langle
f_k v_k |W(k)|f_k v'_k\rangle$ Gaussian and distributed according
to Eq.~(\ref{Wk-mom}). The embedding of this interaction into a
space of different dimension is accomplished by projecting the
product states $|f_k v_k \, \rangle |f_s v_s \, \rangle$ onto the
subspace of states which transform according to the irreducible
representation $D^{f_m}(g)$ contained in the direct product 
$D^{f_k}(g) \times D^{f_s}(g)$. We denote the associated projection
operator by $I(f_m)$. The ensemble is then defined in terms of the
Hamiltonian
\begin{equation}
H(k) = I(f_m) W(k) I(f_m) \; .
\label{Hk-G}
\end{equation}
Eq.~(\ref{Hk-G}) comprises the essence of the group-theoretical 
extension of the idea of an embedded ensemble. This extension is 
independent of the existence of Fermions and Bosons. It relies only 
on group--theoretical concepts.

Using the standard composition formula
\begin{equation}
|r_m f_m v_m \rangle
= \sum_{v_k v_s} | f_k v_k \, \rangle | f_s v_s \, \rangle
C_{f_k v_k f_s v_s}^{r_m f_m v_m} \; ,
\label{cfp-G}
\end{equation}
we find for the matrix elements of $H(k)$
\begin{equation}
H_{r_m v_m , r'_m v'_m}(k)
= \sum_{v_k v'_k v_s}
( C_{ f_k v_k f_s v_s }^{ r_m f_m v_m } )^\ast \,
C_{ f_k v'_k f_s v_s }^{ r'_m f_m v'_m } \,
W_{v_k v'_k}(k) \; .
\label{Hk-cfp-G}
\end{equation}
Here $ C_{ f_k v_k f_s v_s }^{ r_m f_m v_m } $ denotes the
Clebsch--Gordan coefficient of the group $G$ for the coupling
$(f_kf_s)f_m$. The multiplicity index $r_m$ distinguishes different
$D^{f_m}(g)$ of the same highest weight $f_m$ which may appear in
the reduction of $D^{f_k}(g) \times D^{f_s}(g)$. The dimension of the
matrices $H(k)$ is $N=\mu N_m$, where $N_m$ denotes the dimension of
$D^{f_m}(g)$ and $\mu$ the multiplicity with which this representation
appears in the reduction. By construction, the ensemble is invariant
under unitary transformations $U \in \mbox{U}(N_k)$ of dimension
$N_k$ of the interaction matrices $W_{v_k v'_k}(k)$.

In analogy to Eq.~(\ref{Hs-cfp}), we can introduce the dual ensemble
\begin{equation}
H(s) = I(f_m) W(s) I(f_m) \; ,
\label{Hs-G}
\end{equation}
with the random GUE interaction $W(s)$ acting now on the second
system $s$. Eqs.~({\ref{Hk-cfp-G}) and (\ref{Wk-mom}) and analogous
equations valid for the Hamiltonian $H(s)$ and the interaction
$W(s)$ imply for the matrices of second moments $A(k)$ and $A(s)$ of
the two dual ensembles the duality relation (see Eq.~(\ref{mom-dua}))
\begin{equation}
\overline{ N_k H_{r_m v_m, r'_m v'_m}(k)
H_{ {\tilde r}_m {\tilde v}_m, {\tilde r}'_m {\tilde v}'_m }(k) }
= \overline{ N_s H_{ r_m v_m, {\tilde r}'_m {\tilde v}'_m}(s)
H_{ {\tilde r}_m {\tilde v}_m, r'_m v'_m }(s)}  \; .
\label{mom-dua-G}
\end{equation}

To derive the eigenvalue expansion of the matrix of second moments,
we expand the interaction $W(s)$ in terms of a complete set of
$N_s^2$ basic interactions $B_s(\kappa) = B_s^+(\kappa)$ normalized
according to $\langle B_s(\kappa) B_s(\kappa') \rangle_s = \delta_{\kappa
\kappa'}$, with $B_s(0)$ denoting again a normalized multiple of the unit
operator. Here, $\langle \, O_s \, \rangle_s $ denotes the trace of
$O_s$ in the Hilbert space of states $|\, f_s v_s \, \rangle$ 
of the second system. 
This yields
\begin{equation}
W(s) = \sum_{\kappa} B_s(\kappa) W_s(\kappa) \, ,
\qquad
\overline{ W_s(\kappa) W_s(\kappa') }
= \frac{\lambda^2}{N_s} \delta_{\kappa \kappa' } \; .
\label{Ws-G}
\end{equation}
Using this expansion and proceeding as in Section~\ref{mom}, we find
the eigenvalue expansion
\begin{eqnarray} &&
A_{r_m v_m {\tilde r}'_m {\tilde v}'_m, {\tilde r}_m {\tilde v}_m 
r'_m v'_m}(k)
= \overline{ H_{r_m v_m, r'_m v'_m}(k)
H_{ {\tilde r}_m {\tilde v}_m, {\tilde r}'_m {\tilde v}'_m } (k) }
\qquad \qquad \qquad
\nonumber \\ && \qquad 
= \frac{\lambda^2}{N_k}\sum_\kappa \langle r_m f_m  v_m|B_s(\kappa)
| {\tilde r}'_m f_m {\tilde v}'_m \rangle
\langle {\tilde r}_m f_m {\tilde v}_m |B_s(\kappa)|r'_m f_m v'_m \rangle .
\label{mom-exp-G}
\end{eqnarray}
Basic properties of the ensemble can be derived from this formula.
The eigenvalue expansion can eventually be simplified by adapting the 
choice of the operators $B_s(\kappa)$ to the group chain 
$\mbox{U}(N_s) \supset D^{f_s}$. We show in Appendix~\ref{Bsb-tra} that the 
traces $ \langle \, B_s(\kappa) \, \rangle_m$ of the operators 
$B_s(\kappa)$ in the Hilbert space of the
composite system vanish for $\kappa \ne 0$. This makes it possible to
use the supersymmetry approach in a meaningful way. The graded
integrals which represent the $n$-point functions of the ensemble
are dominated by the saddle points and/or saddle--point manifolds
analogous to those discussed in the preceding section.

(ii) We may consider EGUE($k$) as a member of a family of ensembles
obtained by a modification of the GUE, the free Gaussian unitary
ensemble in $N$ dimensions. Eq.~(\ref{Hk-exp}) 
shows that the GUE in $N$ dimensions, GUE = EGUE($m$), may be written as
\begin{equation}
H_{v_m v'_m}
= \sum_{b w_b} \langle \, m v_m |B_m(bw_b)|m v'_m \, \rangle  
W(b w_b) \; ,
\label{H-GUE}
\end{equation}
with the sum over $b$ running over all $b=0, \ldots, m$ and all 
corresponding $w_b$, and
with $W(b w_b)$ Gaussian distributed with mean value zero and second
moment
\begin{equation}
\quad \overline{ W(b w_b) W(b' w'_{b'}) }
= \frac{\lambda^2}{N} \delta_{bb'} \delta_{ w_b w'_{b'} } \, .
\label{H-GUE1}
\end{equation}
When we restrict the sum over $b$ to $b$ not larger than $k$ and
renormalize the matrix elements by the $b$--dependent factors $K(b) =
\sqrt{N/N_k} \langle \, m || B_k(b) || m \, \rangle$, the free
Gaussian ensemble turns into EGUE($k$). Other restrictions of the
sum over $b$ and other renormalizations $K(b)$ of the matrix elements
will create other ``modified GUE ensembles'' of $N \times N$ matrices  
\begin{equation}
{\cal H}_{v_m v'_m}
= {\sum_{b w_b}}' 
\langle \, m v_m |B_m(bw_b)|m v'_m \, \rangle K(b) W(b w_b) \; .
\label{H-GUEA}
\end{equation} 
The prime at the summation symbol indicates the restriction
of the sum over $b$. By construction, the resulting modified ensembles are 
all invariant under the SU($l$) transformation 
${\cal H} \rightarrow D^{f_m}(u) {\cal H} [ D^{f_m}(u) ]^+$.
On expressing  the matrix elements $\langle \, m v_m |B_m(bw_b)
|m v'_m \rangle$ in terms of SU($l$) Clebsch-Gordan coefficients 
and performing the appropriate SU($l$) recoupling, we find for 
the matrix of moments of the modified ensemble the eigenvalue expansion
\begin{equation}
A_{v_m {\tilde v'}_m, {\tilde v}_m v'm}
= \overline{ {\cal H}_{v_m v'_m} 
{\cal H}_{ {\tilde v}_m {\tilde v}'_m } }
= \sum_{b w_b} C_{v_m {\tilde v}'_m}^{g_b w_b} 
\Lambda^b C_{ {\tilde v}_m v'_m }^{g_b w_b} \; .
\label{new55} 
\end{equation}
Here, $C_{v_m v'_m}^{g_b w_b}$ denote the SU($l$) Clebsch-Gordan 
coefficients Eq.~(\ref{Labk-Cbv}), and $\Lambda^b$ the eigenvalues 
\begin{equation}
\Lambda^b = \frac{ \lambda^2 }{N}
{\sum_{\tilde b}}' \sqrt{ \frac{ M_{\tilde b} }{M_b} } K^2(\tilde b)
\langle ( (f_m {\bar f_m} ) g_b (f_m {\bar f}_m) g_b ) 0 |
( (f_m {\bar f_m} ) g_{\tilde b} (f_m {\bar f}_m) g_{\tilde b} ) 0
\rangle \; .
\label{new77}
\end{equation}
The angular bracket expression stands for the SU($l$) recoupling
coefficient for the indicated change of the coupling scheme, with 0
denoting the highest weight of the identity representation.
For $\Lambda^b$ positive, the properties of the modified ensemble thus 
can be studied by applying Efetov's supersymmetric averaging technique
in the same way as for EGUE($k$).

The cases mentioned under (i) and (ii) are very instructive. They
show that our group--theoretical approach to EGUE($k$) is very
general. Morerover, the results obtained for EGUE($k$) are typical
for a wide class of embedded ensembles defined in terms of a general compact
group $G$.

\subsection *{Acknowledgments}
Z. P. thanks the members of the Max-Planck-Institut f\"ur Kernphysik
in Heidelberg for their hospitality and support, and acknowledges 
support by the grant agency GACR in Prague. He is also grateful to 
T. Juza for many stimulating discussions and suggestions.

\appendix

\section{The composition formula for $A^+(m v_m)$}
\label{Anv}

The single--particle creation operators $a_i^+$ transform according
to the SU($l$) representation $D^{f_1}(u)$. The reduction of the
direct product $D^{f_{j-1}}(u) \times D^{f_1}(u)$ contains the
representation $D^{f_j}(u)$ just once. Therefore, the creation
operators $A^+(m v_m)$ can be written as
\begin{equation}
A^+(m v_m)
= \frac{1}{\sqrt{m!}} \sum_{i_1 \ldots i_m}a_{i_1}^+ \ldots a_{i_m}^+
C_{i_1 \ldots i_m}^{f_m v_m} \; ,
\label{Anv-exp}
\end{equation}
where  $C_{i_1 \ldots i_m}^{f_m v_m}$ stands for the coupling
coefficient
\begin{equation}
C_{i_1 \ldots i_m}^{f_m v_m}
= \sum_{v_2 v_3 \ldots v_{m-1}} C_{f_1 i_1 f_1 i_2}^{f_2 v_2}
C_{f_2 v_2 f_1 i_3}^{f_3 v_3}
\ldots
C_{f_{m-1} v_{m-1} f_1 i_{m}}^{f_m v_m} \; .
\label{Ci-fv}
\end{equation}
This coefficient is given in terms of products of the SU($l$)
Clebsch--Gordan coefficients $C_{f_{j-1} v_{j-1} f_1 i_j}^{f_j v_j}$
for the coupling $(f_{j-1}f_1)f_j$, with $j=2, \ldots , m$. In the
Fermionic (Bosonic) case, the coefficients $C_{i_1 \ldots i_m}^{f_m
v_m}$ are totally antisymmetric (symmetric) functions of $i_1,
\ldots, i_m$. The factor $(m!)^{-1/2}$ takes care of the proper
normalization $ \langle \, | A(m v_m) A^+(m v_m)| \, \rangle =
\langle \, m v_m| m v_m \,\rangle = 1$: The total number of
contractions contributing to the norm is equal to $m!$. This
statement applies equally to Fermions and to Bosons.

Indicating the coupling scheme explicitly, we can rewrite
Eq.~(\ref{Anv-exp}) in the symbolic form
\begin{equation}
A^+(m v_m) = \frac{1}{\sqrt{m!}} ( (a^+ a^+)^{f_2} \ldots
a^+)_{v_m}^{f_m} \; .
\label{Anv-sym}
\end{equation}
Modifying the coupling scheme we find
\begin{equation}
( (a^+ a^+)^{f_2} \ldots a^+)_{v_m}^{f_m}
= \sum_{v_k v_s} \,
( (a^+ a^+)^{f_2} \ldots a^+)_{v_k}^{f_k} \,
( (a^+ a^+)^{f_2} \ldots a^+)_{v_s}^{f_s} \,
C_{f_k v_k f_s v_s}^{f_m v_m} \; ,
\label{akas-an}
\end{equation}
where $C_{f_k v_k f_s v_s}^{f_m v_m}$ are the SU($l$)
Clebsch--Gordan coefficients for the coupling $(f_k f_s)f_m$.
Substituting this relation into Eq.~(\ref{Anv-sym}) and introducing
the creation operators $A^+(k v_k)$ and $A^+(s v_s)$, we obtain
\begin{equation}
A^+(m v_m) = \sqrt{ \frac{ k! s!}{m!} }
\sum_{v_k v_s} \, A^+(k v_k) A^+(s v_s) \,
C_{f_k v_k f_s v_s}^{f_m v_m} \; .
\label{com-for}
\end{equation}
This is the composition formula which underlies Eq.~(\ref{cfp}).

Taking into account that, due to the anticommutativity (commutativity)
of the Fermionic (Bosonic) operators $a_i^+$, the products
$A^+(k v_k)A^+(s v_s)$ can be coupled only to operators belonging to
the irreducible tableaux $f_m$, i.e.,
\begin{equation}
\sum_{v_k v_s}A^+(k v_k) A^+(s v_s) C_{f_k v_k f_s v_s}^{fv} = 0
\qquad \mbox{for} \quad f \ne f_m \; ,
\label{new1}
\end{equation}
and making use of orthonormality of Clebsch-Gordan coefficients,
we find that the composition formula can be inverted,
\begin{equation}
A^+(k v_k) A^+(s v_s) 
= { m \choose k}^{1/2} \sum_{v_m} A^+(m v_m) 
( C_{f_k v_k f_s v_s}^{f_m v_m})^* \; .
\label{new2} 
\end{equation}
>From this equation it follows that the matrix elements of the 
creation operators $A^+(k v_k)$ have the form 
\begin{equation}
\langle \, m v_m | A^+(k v_k) | s v_s \rangle 
= {m \choose k}^{1/2} ( C_{f_k v_k f_s v_s}^{f_m v_m} )^* \; .
\label{new3}
\end{equation}
Substituting this equation in
\begin{equation}
\langle \, m v_m |W(k)| m v'_m \, \rangle
= \sum_{v_k v'_k v_s} \langle \, m v_m | A^+(k v_k) | s v_s \, \rangle
\langle \, s v_s | A(k v'_k) | m v'_m \, \rangle W_{v_k v'_k}(k) \;
\label{new4}
\end{equation}
yields the formula Eq. (\ref{Hk-cfp}).

\section{The matrices $\Delta^k(U)$}
\label{DkU-pro}

Multiplying Eq.~(\ref{DkU-def}) by $\langle \, k v'_k |B_k(\kappa'')| k v_k \,
\rangle$ and summing it over $v_k$ and $v'_k$ with the help of the
orthonormality relation~(\ref{Bkb-pro}), we arrive at the following explicit
expression for $\Delta_{\kappa' \kappa}^k(U)$ :
\begin{equation}
\Delta_{\kappa' \kappa}^k(U)
= \sum_{v_k {\tilde v}_k v'_k {\tilde v'}_k}
U_{v_k {\tilde v}_k}
\langle \, k {\tilde v}_k |B_k(\kappa)|k {\tilde v}'_k \, \rangle
(U^+)_{ {\tilde v}'_k v'_k}
\langle \, k v'_k |B_k(\kappa')|k v_k \, \rangle \; .
\label{DkU}
\end{equation} 
We substitute $\langle \, k v_k| B_k(0) |k v'_k \, \rangle 
= \delta_{v_k v'_k} (N_k)^{-1/2}$, use Eq.~(\ref{DkU}), and find that 
\begin{equation}
\Delta_{0\kappa}^k(U)=\Delta_{\kappa 0}^k(U)= \delta_{\kappa 0} \; .
\label{new11}
\end{equation}
The representation $\Delta^k(U)$ thus has a fully reduced form:
Under the unitary transformations $U \in \mbox{U}(N_k)$, 
the matrix of $B_k(0)$ remains invariant whereas the matrices of
the $B_k(\kappa)$ with $b>0$ transform like the components of an
irreducible U($N_k$) tensor of dimension $N_k^2-1$ and of Young tableau
$(10^{N_k-3}1)$. For $U=D^{f_k}(u)$, from Eq.~(\ref{TBT}) it follows that
\begin{eqnarray}
&& \sum_{ {\tilde v}_k {\tilde v'}_k} D_{v_k {\tilde v}_k}^{f_k}(u)
\langle \, k {\tilde v}_k |B_k(b w_b)|k {\tilde v}'_k \, \rangle
([D^{f_k}(u)]^+)_{ {\tilde v}'_k v'_k}
\nonumber \\ && \qquad \qquad \qquad
= \sum_{w'_b} D_{w'_b w_b}^{g_b}(u)
\langle \, k v_k |B_k(b w'_b)|k v'_k \, \rangle \; .
\label{DBD-DB}
\end{eqnarray}
For $U = D^{f_k}(u)$, the matrices $\Delta^k(U)$ simplify to
$\Delta_{\kappa' \kappa}^k(U) = \delta_{bb'}D_{w'_{b'} w_b}^{g_b}(u)$.

\section{The reduced matrix elements $\langle \, m || B_s(b) ||
m \, \rangle $}
\label{red-ele}

To calculate the squares of the reduced matrix elements, we apply
the approach by Mon and French~\cite{mon75} (cf. also Refs.
\cite{par78,won86}). At the end of Section~\ref{for} it was shown 
that the value of $\langle \, m || B_s (b) || m  \, \rangle^2$ 
can be obtained from the norm $\langle\, S_s^+(b) S_s(b) \, 
\rangle_m$ of a suitable $s$-particle operator $S_s(b)$ of bodyness 
$b$. Any such operator can be written as the product
\begin{equation}
S_s(b) = { {\hat N - b} \choose {s-b} } S_b(b) \;
\label{Ss-Sb}
\end{equation}
of the SU($l$) invariant polynomial of the particle number operator
$\hat N$ given by the first factor and of a $b$-particle operator
of bodyness $b$ denoted by $S_b(b)$. The form of the polynomial is
uniquely determined by the particle ranks of $S_s(b)$ and $S_b(b)$:
Since $S_s(b)$ is an $s$-particle operator, the polynomial has to
vanish when acting on the states of $t=b, \ldots , s-1$ particles.
This reduces the choice of $S_s(b)$ to the choice of $S_b(b)$.
The $b$-particle operators of bodyness $b$ change the state of
$b$ particles and cannot be simulated by the operators of lower
particle rank. For Fermions, a suitable choice then is $ S_b(b) =
\xi a_1^+ \ldots a_b^+a_{b+1} \ldots a_{2b}$ while for Bosons, we
use $S_b(b) = \xi ( a_1^+)^b ( a_2 )^b/b! \;$. Here, $\xi$ denotes
a normalization factor. This yields
\begin{equation}
\langle \, S_b^+(b) S_b(b) \, \rangle_m = \xi^2{ {l-2b} \choose
{m-b} } \; , \; \; \; \langle \, S_b^+(b) S_b(b) \, \rangle_m =
\xi^2{ {l + m + b - 1} \choose {m-b} } \;
\label{norb-fb}
\end{equation}
for Fermions and Bosons, respectively. We substitute this result in
\begin{equation}
\langle \, S_s^+(b) S_s(b) \, \rangle_m
= { {m-b} \choose {s-b} }^2
\langle \, S_b^+(b) S_b(b) \, \rangle_m
\label{nors-norb}
\end{equation}
and fix the value of $\xi$ from the normalization condition $\langle
\, S_s^+(b) S_s(b) \, \rangle_s = 1$. This yields
\begin{eqnarray}
&&\langle \, m || B_s (b) || m \, \rangle^2
= \langle \, S_s^+(b) S_s(b) \, \rangle_m
\nonumber \\ 
&& \qquad = { {m - b} \choose k }{ {l - m  + k - b} \choose k } \;
\; \; \; {\rm (Fermions)} \nonumber \\
&& \qquad = { {m - b} \choose k }{ {l + m + b - 1} \choose k } \;
\; \; \; {\rm (Bosons)} \ ,
\label{red-fb}
\end{eqnarray}
in keeping with Eqs.~(\ref{Labk-fer}) and (\ref{Labk-bos}).

\section{The traces $\langle \, B_s(\kappa) \, \rangle_m$ }
\label{Bsb-tra}

Using the composition formula~(\ref{cfp-G}) we find that
\begin{equation}
\langle \, B_s(\kappa) \, \rangle_m
= \sum_{ r_m v_m v_k v_s v'_s }
( C_{f_k v_k f_s v_s}^{r_m f_m v_m} )^\ast
C_{f_k v_k f_s v'_s}^{r_m f_m v_m}
\langle \, f_s v_s |B_s(\kappa)| f_s v'_s \, \rangle \; .
\label{tra-CGC}
\end{equation}
We express the Clebsch--Gordan coefficients with the help of the
symmetry relation (cf. Refs.~\cite{wyb74, but75})
\begin{equation}
C_{ f_k v_k f_s v_s }^{ r_m f_m v_m }
= \sqrt{ \frac{ N_m }{ N_s } }
\sum_{ {\bar v}_k r_s}
( C_{ {\bar f}_k {\bar v}_k f_m v_m}^{r_s f_s v_s} )^\ast
U_{v_k {\bar v}_k }^{ (f_k) } U_{r_s r_m}^
{(f_k f_s f_m) } \; .
\label{CGC-sym}
\end{equation}
Here, ${\bar f}_k$ denotes the highest weight of the irreducible 
representation conjugate to $D^{ f_k }(g)$; ${\bar v}_k$  
the basis state of this representation; 
$C_{ {\bar f}_k {\bar v}_k f_m v_m}^{r_s f_s v_s}$
the Clebsch-Gordan coefficient for the coupling 
$({\bar f}_k f_m)f_s$; $r_s$ the multiplicity index
for this coupling; $U_{ v_k {\bar v}_k }^{ (f_k) }$ 
and $U_{r_s r_m}^{(f_k f_s f_m)}$ the  unitary matrices
\begin{equation}
U_{ v_k {\bar v}_k }^{(f_k )}
= \sqrt{N_k} C_{ f_k v_k {\bar f}_k {\bar v}_k}^{00} \; ,
\label{S1-mat}
\end{equation}
\begin{equation}
U_{r_s r_m}^{(f_k f_s f_m)}
= \sqrt{ \frac{1}{N_sN_m} }
\sum_{v_k {\bar v}_k v_s v_m}
C_{ {\bar f}_k {\bar v}_k f_m v_m }^{ r_s f_s v_s }
C_{ f_k v_k f_s v_s }^{ r_m f_m v_m }
( U_{v_k {\bar v}_k }^{(f_k)} )^\ast \; .
\label{S3-mat}
\end{equation}
In Eq.~(\ref{S1-mat}), (00) denotes the $(fv)$ labels of the 
identity representation. Substituting Eq.~(\ref{CGC-sym})
and performing the sum yields 
\begin{equation}
\langle \, B_s(\kappa) \, \rangle_m
= \frac{N}{N_s} \langle \, B_s(\kappa) \, \rangle_s
= \delta_{\kappa 0} \frac{N}{ \sqrt{N_s} } \; .
\label{tra-res}
\end{equation}
The traces are thus nonzero only for $\kappa = 0$.

\end{document}